\documentclass[prx, superscriptaddress,aps, english, twocolumn, hyperref, nofootinbib]{revtex4-1}

\usepackage{graphicx}

\usepackage{amssymb, amsmath, color, rotating}

\begin{document}

\title{Competing ground states of strongly correlated bosons in the Harper-Hofstadter-Mott model}

\author{Stefan S. Natu}

\email{snatu@umd.edu}

\affiliation{Condensed Matter Theory Center and Joint Quantum Institute, Department of Physics, University of Maryland, College Park, Maryland 20742-4111 USA}

\author{Erich J. Mueller}

\affiliation{Laboratory of Atomic and Solid State Physics, Cornell University, Ithaca 14853 USA}

\author{S. Das Sarma}

\affiliation{Condensed Matter Theory Center and Joint Quantum Institute, Department of Physics, University of Maryland, College Park, Maryland 20742-4111 USA}

\begin{abstract}

Using an efficient cluster approach, we study the physics of two-dimensional lattice bosons in a strong magnetic field in the regime where the tunneling is much weaker than the on-site interaction strength. We study both dilute, hard core bosons at filling factors much smaller than unity occupation per site, and the physics in the vicinity of the superfluid-Mott lobes as the density is tuned away from unity.  For hardcore bosons, we carry out extensive numerics for a fixed flux per plaquette $\phi=1/5$ and $\phi = 1/3$.  At large flux, the lowest energy state is a strongly correlated superfluid, analogous to He-$4$, in which the order parameter is dramatically suppressed, but non-zero. At filling factors $\nu=1/2,1$, we find competing incompressible states which are metastable.  These appear to be commensurate density wave states. For small flux, the situation is reversed, and the ground state at $\nu = 1/2$ is an incompressible density-wave solid. Here, we find a metastable lattice supersolid phase, where superfluidity and density-wave order coexist. We then perform careful numerical studies of the physics near the vicinity of the Mott lobes for $\phi = 1/2$ and $\phi = 1/4$. At $\phi = 1/2$, the superfluid ground state has commensurate density-wave order. At $\phi = 1/4$, incompressible phases appear outside the Mott lobes at densities $n = 1.125$ and $n = 1.25$, corresponding to filling fractions $\nu = 1/2$ and $1$ respectively. These phases, which are absent in single-site mean-field theory are metastable, and have slightly higher energy than the superfluid, but the energy difference between them shrinks rapidly with increasing cluster size, suggestive of an incompressible ground state. We thus explore the interplay between Mott physics, magnetic Landau levels, and superfluidity, finding a rich phase diagram of competing compressible and incompressible states.

%We capture local quantum fluctuations on small clusters exactly, and couple the clusters using mean-fields. %The cluster technique, we use can be readily generalized to different lattice geometries, higher densities, long range hoppings and interactions, and offers a computationally inexpensive method to explore correlated bosonic phases.

%Using an efficient cluster approach, we study the physics of two-dimensional lattice bosons in a strong magnetic field in the regime where the tunneling is much weaker than the on-site interaction strength.  

\end{abstract}
\maketitle

\section{Introduction}

Ever since the discovery of the quantum Hall effects \cite{Laughlin83, Klitzing80, Tsui82}, attention has focussed on understanding how the underlying lattice affects the properties of electrons in a magnetic field. Following the work of Harper \cite{Harper55}, Hofstadter showed that the Bloch bands fragment into smaller bands that acquire an intricate self-similar spectrum as a function of magnetic field, known as the Hofstadter butterfly \cite{Hofstadter76}. The integer Quantum Hall effect, which is a single-particle phenomenon, not only survives in this fractal spectrum, but becomes richer \cite{TKNN82}.   A similar richness is expected of the strongly interacting system.  Experimental studies of this strongly interacting lattice problem in a strong magnetic field (which we call the Harper-Hofstadter-Mott model), however, have been challenging in two-dimensional electronic systems \cite{Melinte04, Feil07, Albrecht01, Geisler04} as the magnetic fields required for the Landau levels to sufficiently alter the structure of the Bloch bands are typically extremely large, and weak disorder, invariably present in all solid state systems, can easily overwhelm the fractal Hofstadter structure. Ultra-cold atomic gases offer a new and promising avenue to realizing this physics by using Raman lasers to imprint a phase on the motion of the atoms as they hop from site to site \cite{Atala14, Aidelsburger13, Miyake13, Aidelsburger15}. Bosons or fermions loaded into such lattices experience an artificial magnetic field, and are thus sensitive to the underlying fractal band structure. Here we study the physics of strongly interacting bosons on a square lattice in a large magnetic field, \textit{i.e.}, we study theoretically the physics of the Harper-Hofstadter-Mott model with three independent energy scales: Bloch bands, Landau levels, and strong correlations.  The interesting (and theoretically challenging) situation of course arises when these energy scales are all comparable, and consequently, perturbative or purely mean-field techniques may not work.

Bosonic analogs of quantum Hall effects  have been theoretically studied since the 1980's \cite{Haldane83}, and the fermion-boson mapping in the context of the continuum fractional quantum Hall effect is well understood in the lowest Landau level \cite{Sarma91}.  Cold atoms appear to be one of the best systems for realizing these
states \cite{Greiner02, Fisher89}, since there are no known bosonic solid state materials suitable for quantum Hall effect experimental exploration.  For example, by rapidly rotating a low density harmonically trapped gas, various authors have predicted that
%in the continuum, it is well known that in the limit of rapid rotation and low boson density,
the ground state will be a correlated liquid such as the $\nu = 1/2$ bosonic Laughlin state or the bosonic Moore-Read Pfaffian state at $\nu = 1$ \cite{Cooper08, Cooper01, Wilkin00, Regnault04}. Here $\nu=N/N_\phi$ is the ratio between the number of particles $N$ and the number of flux quanta $N_\phi$.  Using exact diagonalization (ED), S\o{}rensen, Hafezi, Demler and Lukin \cite{Sorensen05, Hafezi07} showed that for weak magnetic fields, the $\nu = 1/2$ Laughlin state survives even in the lattice, but is destroyed at larger magnetic fields. M\"oller and Cooper \cite{Moller09} have argued that composite fermion-like states occur at filling fractions other than $1/2$, but the overlap of the exact ground state with these trial wave functions diminishes rapidly with increasing particle number or flux. Using variational Monte Carlo and single site mean-field theory, Umucalilar, Oktel and Mueller \cite{Onur07, Onur09} have shown that in the vicinity of the Mott lobes, away from integer filling, excess particles form a quantum Hall state above a uniform Mott insulating background.

%Although the bosonic quantum Hall effect has been known theoretically for a long time \cite{Haldane83}, the realization of the superfluid-Mott transition of neutral atoms in an optical lattice \cite{Greiner02, Fisher89} implied that the bosonic quantum Hall effect could potentially be explored in an experimental setting. In the continuum, it is well known that in the limit of rapid rotation and low boson density, the ground state is a correlated liquid such as the $\nu = 1/2$ Laughlin state, or the Pfaffian state at $\nu = 1$ \cite{Cooper08, Cooper01, Wilkin00, Regnault04}. Using exact diagonalization (ED), S\o{}rensen, Hafezi, Demler and Lukin \cite{Sorensen05, Hafezi07} showed that for weak magnetic fields, the $\nu = 1/2$ Laughlin state survives even in the lattice, but is destroyed at larger magnetic fields. M\"oller and Cooper \cite{Moller09} have argued that composite fermion-like states occur at filling fractions other than $1/2$, but the overlap of the exact ground state with these trial wave functions diminishes rapidly with increasing particle number or flux. Using variational Monte Carlo and single site mean-field theory, Umucalilar, Oktel and Mueller \cite{Onur07, Onur09} have shown that in the vicinity of the Mott lobes, away from integer filling, excess particles form a quantum Hall state above a uniform Mott background. 

Here we employ a new type of cluster variational method to nonperturbatively study the physics of interacting lattice bosons in a large magnetic field. We incorporate short range correlations by diagonalizing small clusters exactly, and capture thermodynamic features by coupling neighboring clusters using mean-fields. In our first set of calculations, we compute the equation of state (EOS) for hardcore bosons in the Harper-Hofstadter-Mott model on a square lattice for flux $\phi = 1/3$ and $\phi = 1/5$, where $\phi$ is the number of flux quanta per site. We show that at flux $\phi = 1/3$, as long as there is less than one particle per site, the ground state is a strongly correlated \textit{superfluid}, characterized by a small but non-vanishing condensate fraction. Precisely at filling fractions $\nu = 1/2$ and $1$, metastable \textit{incompressible} phases compete with superfluidity. These appear to be commensurate checkerboard solids with density wave order. For smaller values of flux, $\phi = 1/5$, the situation is reversed, and the ground state found by this method at $\nu = 1/2$  is incompressible, and supports stripe order. The metastable superfluid phase at $\nu = 1/2$, contains density wave order, and can be considered a supersolid \cite{Chan04}.

The EOS has the advantage that unlike spectral gaps or braiding properties, it is directly measurable in cold atom experiments, and provides insight into thermodynamic quantities, such as the compressibility. The compressibility is the analog of the longitudinal resistance measured in electronic systems. A vanishing longitudinal resistance, or (the associated) quantized Hall conductance is the hallmark of a fractional quantum Hall liquid. In our approximation, the clusters become disconnected when the condensate fraction vanishes.  Under those circumstances, finite size effects become significant.  Thus it is difficult for us to reliably distinguish between charge density wave states and incompressible quantum Hall fluids.  We believe that the sequence of compressible and incompressible states is robust, despite this ambiguity in the identity of the incompressible phase. Within our approach, we find no direct evidence for a Laughlin state at $\nu = 1/2$, or any other fractional quantum Hall liquid at $\nu = 1$, for any flux. 

In our second set of calculations, we go beyond the hard core limit and explore the physics near the superfluid-Mott transition as the density is tuned away from unity filling, focussing on the experimentally realized values of flux \cite{Aidelsburger13, Miyake13, Aidelsburger15}, $\phi = 1/4$ and $\phi = 1/2$. For $\phi = 1/4$, and sufficiently large cluster sizes, we find an incompressible state at $n = 1.125$ corresponding to $\nu = 1/2$ for $4\times2$ clusters. This state, which is absent in single-site mean-field theory, has \textit{higher} energy than the superfluid at the same density, but the energy difference between the two decreases rapidly with increasing cluster size (and the possibility that this incompressible state is indeed the ground state in the thermodynamic limit cannot be ruled out). Within our theory, this non-condensed state does not correspond to a Laughlin liquid -- again this is related to finite size effects from diagonalizing small clusters. At $\phi = 1/2$, we improve the single site mean-field results of Umucalilar and Oktel \cite{Onur07} to determine the superfluid-Mott phase boundary at unity density, and show that the superfluid ground state has robust density-wave order super-imposed on it (i.e. is is a type of a supersolid), directly measurable in experiments.

Although we focus here on the Harper-Hofstadter-Mott model, our approach can be readily extended to other tight binding lattice models with topologically non-trivial flat bands \cite{Kai11, Neupert11, Tang11}. By flattening the band, interaction effects are enhanced, allowing the possibility for fractional quantum Hall states in the absence of Landau levels \cite{Haldane88}, so called fractional Chern insulators \cite{Moller09, Sheng11, Kapit10, Bernevig11, Sid13, Gu11}. Such band structures are actively being explored experimentally in ultra-cold atomic systems as well as photonic lattices and graphene superlattices \cite{Jo12, Aidelsburger15, Taie15, Dean13, Khanikaev13} and future applications of our numerical method should shed light on the physics of bosonic fractional Chern insulators in flat band systems. %(see Ref.~\cite{Sid13} and references therein). Such band structures have been recently realized in 

This paper is organized as follows: In Sec.~II we outline the cluster method we use, its salient features and limitations in detail. In Sec.~III, we study dilute hardcore bosons, where the density is much smaller than unity filling per site. In Sec.~IV, we study the physics near the superfluid-Mott phase boundary, for high densities at and near unity filling per site. Wherever possible, we make comparisons with the earlier literature on this subject. We discuss the experimental signatures in Sec.~V and present conclusions in Sec.~VI.

\section{Cluster Mean-field Method}

In the grand canonical ensemble, the Hamiltonian for the Harper-Hofstadter-Mott model on a square lattice of lattice spacing $a$ is:
\begin{align}\label{hofham}
{\cal{H}} &= -t \sum_{\langle jk \rangle} (e^{i{\cal{A}}_{jk}}a^{\dagger}_{j}a_{k} + \text{h.c}) %\\\nonumber  &
+ \sum_{j} \Big[\frac{U}{2}n_{j}(n_{j}-1) - \mu~n_{j} \Big]
\end{align}
where $\langle jk \rangle$ denotes nearest neighboring sites $j = (j_{x}, j_{y})$, $k = (k_{x}, k_{y})$; $a_{j}$ denotes the bosonic annihilation operator on site $j$, and $n_{j} = a^{\dagger}_{j}a_{j}$ is the density operator on site $j$. Here $t$, $U$ denote the hopping and on-site interaction respectively, and $\mu$ is the chemical potential, which we assume to be spatially uniform. For numerical convenience, we work in the Landau gauge, ${\cal{A}}_{jk} = 0$ on horizontal bonds (ie. when ${\bf r}_k={\bf r}_j\pm a \hat{x}$).  On vertical bonds
${\cal{A}}_{jk} = \pm2\pi\phi j_{x}$, if ${\bf r}_k={\bf r}_j\pm a\hat{y}$. This corresponds to a spatially uniform magnetic field with dimensionless flux $\phi$ through each plaquette. All physical quantities are independent of our gauge choice. %Here we study these two cases in detail. %The flux corresponds to a spatially varying vector potential $\vec{\cal{A}}_{j} = (0, 2\pi \phi j_{x})$. 

In the ultra-cold atom context, the single-particle Hamiltonian (\ref{hofham}) was recently realized at MIT and Munich \cite{Miyake13, Aidelsburger13}, by first introducing a linear potential gradient, which turns off tunneling between neighboring sites in the $\hat y$ direction, and reintroducing the hopping using Raman beams. The lasers impart a momentum kick $k_{R}$ to the bosonic $^{87}$Rb atoms as they hop, proportional to the laser wavelength. The flux $\phi = k_{R}/k_{L}$, where $k_{L} = 1/a$ \cite{Dalibard11}. The original experiments realized the $\phi = 1/2$ limit, where the hopping is real, but alternates in sign from site to site. Aidelsburger \textit{et al.} \cite{Aidelsburger15} have recently extended this to $\phi = 1/4$, and directly measured the Chern number associated with the Hofstadter bands. 
%Recently, $\phi = 1/4$ has also been realized experimentally in quasi-$1$ dimensional ladders, and in a square lattice by Atala \textit{et al.} %\cite{Atala14}. 

We consider a two-dimensional system of size $K \times L$, which we divide into $w$ clusters $({\cal{C}})$ of size $M \times N$. As we detail below, we exactly treat the physics of each cluster, but treat the influence of one cluster on another via mean-fields.  This approach is variational, and yields upper bounds to the energy.  In the special case where $M=N=1$, it reduces to the Gutzwiller mean-field theory \cite{Rokhsar91}.  In the case where $M=K$ and $N=L$ it is exact.

 We write the local cluster wave-function in a Fock basis as $\Psi_{c} = \sum_{m_{1}...m_{MN}}|m_{1}...m_{MN}\rangle$, where $m_{i}$ denotes the local occupation on site $i$, which ranges from $0$ to $k-1$. We decompose the hopping Hamiltonian into two parts: an exact part defined on all internal links of the cluster ($\partial {\cal{C}}$) and a mean-field part defined only for sites on the boundary of the cluster. The mean-fields $\langle\alpha_{k}\rangle$ are given by $\langle\alpha_{k}\rangle = \sum_{j \in \not {\cal{C}}}\langle a_{j} \rangle$, where the sum is over all nearest neighbors $j$, not in ${\cal{C}}$.

The Hamiltonian of a single cluster then reads: 
\begin{align}\label{clustham}
{\cal{H}}_{{\cal{C}}} &= -t\sum_{\langle jk \rangle \in {\cal{C}}} (e^{i{\cal{A}}_{jk}}a^{\dagger}_{j}a_{k} + \text{h.c}) + \sum_{j \in {\cal{C}}}\frac{U}{2}n_{j}(n_{j}-1) - \mu n_{j} \\\nonumber& -t\sum_{k \in {\partial {\cal{C}}}} (e^{i{\cal{A}}_{jk}}\langle\alpha_{k}\rangle^{*} a_{k}  + \text{h.c})
\end{align}
By construction, Eq.~\ref{hofham} reduces to a sum over cluster Hamiltonians ${\cal{H}} = \sum_{{\cal{C}}}{\cal{H}}_{{\cal{C}}}$, coupled by local boundary mean-fields. Throughout, we implement periodic boundary conditions on the full $K \times L$ system. The clusters become independent of one another in the limit of vanishing mean-fields. 

 We minimize the variational energy $E = \sum^{w}_{{\cal{C}}=1}E_{\cal{C}}$, where $E_{\cal{C}} = \langle\Psi_{\cal{C}}|{\cal{H}}_{{\cal{C}}} |\Psi_{\cal{C}}\rangle$ and $w = KL/MN$ by numerically solving for the variational mean-field parameters using an iterative procedure. Starting with a trial set of mean-fields defined on the entire system, we obtain the ground state of each cluster sequentially, updating the mean-fields with their new values obtained from previous clusters. In this manner, we sequentially step through all the clusters and obtain a variational energy after one iteration. We repeat this procedure until the variational energy after successive iterations is stationary up to a desired tolerance $\epsilon/U \leq 10^{-5}$ (or $\epsilon/t \leq 10^{-5}$ in the hard core limit). Lowering the tolerance further does not change the ground state obtained. Typically, most solutions converge after less than $20$ iterations. Convergence is somewhat slower for flux values different from $\phi = 1/2$ as the mean-fields acquire complex values. 

This is a highly nonlinear problem, and our algorithm generally has several basins of attraction, corresponding to local minima of the energy. This is particularly true when the superfluid phase has vortices.  The vortices become pinned to the lattice and different spatial configurations of the vortices have slightly different energies. While it is an interesting intellectual activity to determine which of these vortex configurations have the lowest energy, in practice it is irrelevant, and the experimentally observed configuration will depend on the details of the state preparation.  Our primary concern here is robust features, such as phase transitions between states with and without superfluid order.  Because all the vortex configurations have similar energy, all of them yield nearly identical phase boundaries.
 We repeat our calculations for a number of different initial mean-fields in order to broadly sample the phase space, and verify this feature.
 %Although we occasionally find other superfluid solutions with lower energy than what we present, these solutions do not converge to the tolerance specified.
 Our iterative procedure is closely related to evolving the cluster mean-field equations in imaginary time \cite{Luhmann13}.

%To explore the physics near the vicinity of the Mott lobes, we focus on the $n= 1$ Mott lobe for simplicity, and restrict our Hilbert space to $k =3$, or $0, 1, 2$ particles per site. 
Throughout, we use periodic boundary conditions on the entire $K \times L$ system. We explore a variety of cluster sizes and %present results for $3\times 2$ and $4\times 2$ clusters at $\phi = 1/2$ and $\phi = 1/4$. By exploring 
a wide range of chemical potentials.  This allows us to study several different filling fractions, unlike previous studies \cite{Onur09, Sorensen05, Hafezi07, Moller09, Moller15}, which tend to focus on a few particular choices of $\nu$ where FQH states can appear. 

In the absence of a magnetic field $\phi = 0$, L\"uhmann showed that the superfluid-Mott phase boundary obtained using relatively modest cluster sizes of $3\times3$ or $4\times3$ with periodic boundary conditions already yields strikingly good agreement with the Quantum Monte Carlo (QMC) results of Sansone \textit{et al.} \cite{Luhmann13, Sansone08}. For non-zero $\phi$, QMC incurs a sign problem because of the complex hopping terms, severely restricting its applicability. The Density Matrix Renormalization group (DMRG) technique has recently been employed to study this problem in quasi-$1$D ladder systems \cite{Atala14, Piraud15, Petrescu13, Petrescu15}, finding evidence for Laughlin states only for sufficiently strong nearest-neighbor interactions. In $2$D, physics near the superfluid-Mott transition in the Hofstadter Hamiltonian has been addressed using single-site mean-field theory \cite{Onur07} and variational Monte Carlo \cite{Onur09}. Although the latter study found evidence for a $\nu =1/2$ Laughlin state at densities slightly larger than one particle per site, it is unclear whether the Laughlin wave-function wins over (\textit{i.e.} has lower energy than) better trial superfluid wave-functions. ED studies for small particle numbers at $\nu =1/2$ find that the ground state overlap with the Laughlin wave function or other composite fermion wave-functions decreases rapidly with increasing flux \cite{Sorensen05, Hafezi07, Moller09}. While mean-field and variational methods often do not allow for correlated states such as strongly interacting superfluids or incompressible solids, ED is restricted to small systems, making it hard to make predictions about the thermodynamical limit. Our hybrid technique allows us to capture quantum correlations within a local area exactly, while also inferring thermodynamic properties such as the compressibility and the superfluid fraction. Thus, our cluster-mean-field technique may be ideally suited to numerically study the complexity of various competing interacting ground states in the Harper-Hofstadter-Mott system-- it is less exact than exact diagonalization, but can be used to study larger systems.

%In the presence of a magnetic field, QMC encounters a sign problem even for bosons, and progress must be made using variational approaches \cite{Onur09, natu-hof}, small system exact diagonalization \cite{Sorensen05, Hafezi07, Moller09, Moller15} or Density Matrix Renormalization group in quasi-$1$D systems \cite{Atala14, Piraud15, Petrescu13, Petrescu15}. The cluster method complements this approaches by combining aspects of exact diagonalization with mean-field theory.%In Ref.~\cite{natu-hof}, we developed this technique to include a magnetic field, and obtained the equation of state of hardcore bosons at flux $\phi = 1/3$ and $1/5$. There we showed that for small cluster sizes, incompressible density wave states appear at $\nu = 1/2$ and $1$, which compete with superfluidity. In the weak magnetic field limit $\phi \rightarrow 0$, we expect these incompressible phases to be connected to continuum Laughlin liquids, but for large field, the ground state is always superfluid. Here we go beyond the hardcore limit, and present results on the superfluid-Mott transition in the presence of a large magnetic field. 

\section{Dilute Limit: Equation of state for hardcore bosons} 

We first study the strongly interacting, but low density limit, where the occupation is much smaller than one particle per site. We consider hard core interactions, $U \rightarrow \infty$, by restricting the Hilbert space to $0$ and $1$ particles per site ($k = 2$). At fixed flux $\phi$, there is a single parameter $\tilde \mu = \mu/t$, which we vary. We fix $\phi$ and obtain the cluster wave-functions through the iterative procedure outlined above, and compute average density $n_{\cal{C}} =  1/(w MN)\sum_{{\cal{C}}}\langle\Psi_{\cal{C}}|n |\Psi_{\cal{C}}\rangle$, where $n = \sum_{i \in {\cal{C}}}n_{i}$ is the cluster density operator, and the average superfluid density $\rho_{{\cal{C}}} = 1/(w MN)\sum_{{\cal{C}}}\sum_{k \in \partial {\cal{C}}}|\langle\alpha_{k}\rangle|^{2}$ as a function of $\tilde\mu$. 

\subsection{$\phi = 1/3$}

\begin{figure}
\begin{picture}(100, 170)
\centering
\put(-60, 90){\includegraphics[scale=0.55]{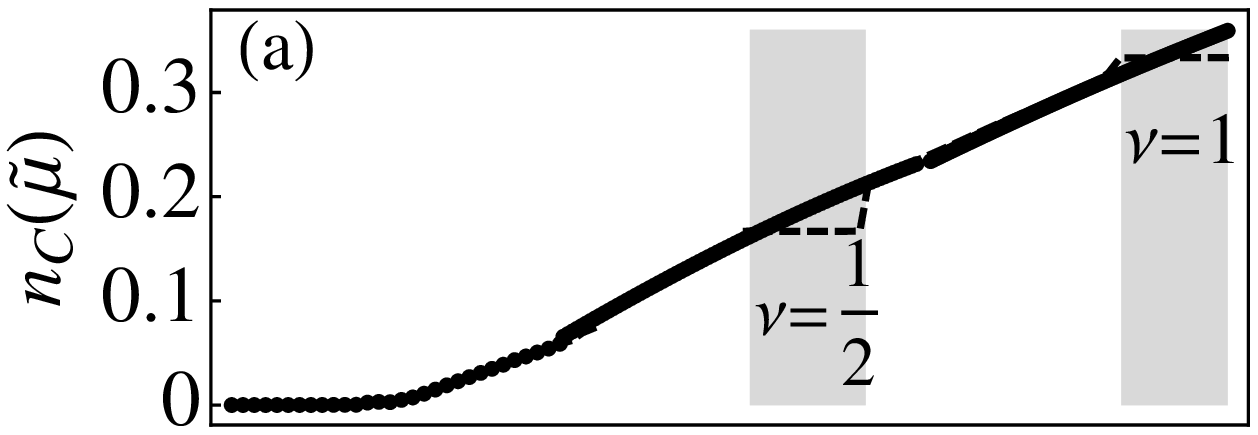}}
\put(-68, -10){\includegraphics[scale=0.573]{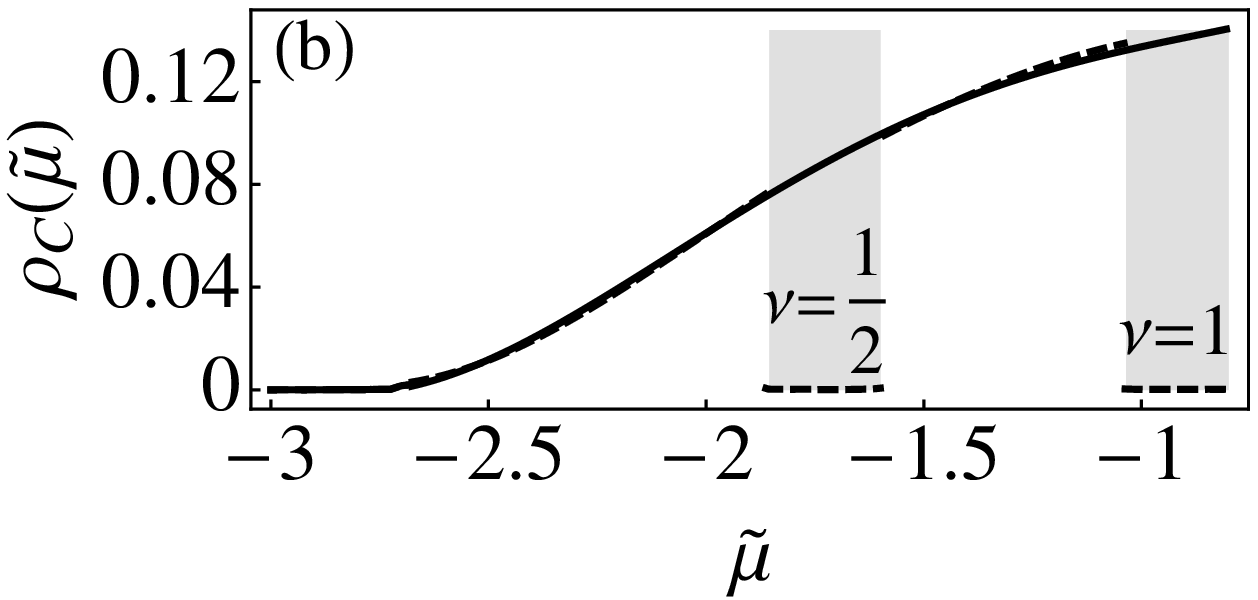}}
%\put(70, 55){\includegraphics[angle=270, origin=c,scale=0.39]{condoverdens.eps}}

\end{picture}
\caption{\label{eosplot} (a) Zero temperature equation of state ($n_{{\cal{C}}}$ vs. $\mu$) of a hardcore Bose gas in an artificial magnetic field. Solid and dashed curves indicate the results from different initial choices of the wave-function. They converge to the same solution except at densities $n_{{\cal{C}}} = 1/6$ and $n_{{\cal{C}}}=1/3$, which correspond to filling factors $\nu = 1/2, 1$ respectively. Incompressible plateaus, marked by the shaded grey regions are observed which have higher energy than the compressible superfluid ground state. (b) Superfluid order parameter $\rho_{{\cal{C}}}$ as a function of $\tilde\mu$, showing that incompressible phases (dashed) are uncondensed. When they differ the energies of the states corresponding to the solid line have lower energy than the dashed.%(c) Superfluid fraction: ratio of superfluid density to cluster density plotted versus cluster density in the ground state. Even at relatively dilute densities of $n_{{\cal{C}}} = 1/3$, the superfluid fraction is suppressed indicative of strong correlations.
}
\end{figure}

We start by considering a single cluster $w=1$, and perform extensive numerics for $\phi = 1/3$. In the Landau gauge, the Hamiltonian becomes periodic over a $3\times 1$ unit cell, and we choose a $M \times N = 4 \times 3$ cluster with periodic boundary conditions. In Fig.~\ref{eosplot}(a), we plot the $\text{T}= 0$ EOS of a dilute, hardcore Bose gas in the Harper-Hofstadter-Mott model. The solid points have \textit{lower} mean-field energy, and point to a compressible ground state, indicated by a non-zero value of $\kappa ^{-1} = \partial n/\partial \mu$. The dashed line corresponds to a different initial condition which has lower value of the condensate order parameter. The dashed solution is identical to the solid points everywhere except at $n= 1/6$ and $n= 1/3$, where we obtain a second \textit{metastable} solution. The vanishing derivative $\kappa ^{-1}$ indicates that this phase is \textit{incompressible}. Furthermore, it corresponds to a non-condensed state at fractional filling with a vanishing superfluid order parameter, as shown in Fig.~\ref{eosplot}(b) (dashed curve). 

We define the superfluid fraction as $\rho_{{\cal{C}}}/n_{{\cal{C}}}$. In the ground state, the superfluid fraction decreases monotonically with density. At low densities, corresponding to large negative $\tilde \mu$, the superfluid fraction is large $\sim 75\%$, consistent with a dilute Bose gas. With increasing chemical potential, the superfluid fraction initially drops rapidly, and then more slowly at higher densities. Surprisingly, even at a relatively dilute density of $n \approx 1/3$, the superfluid fraction in the ground state is suppressed ($\sim 35\%$). We conclude that for flux $\phi = 1/3$, the ground state of hardcore bosons in the Harper-Hofstadter-Mott model is a strongly correlated superfluid, with a small superfluid fraction, analogous to He$-4$ \cite{McMillan65}. This result
 is consistent with ED studies, which find very little overlap between the exact ground state and trial FQH wave-functions \cite{Sorensen05, Moller09}.

%For $\nu = 1/2$, the exact ground state for $N = 4$ was first obtained by Sorensen \textit{et al.}, who found that at $\phi = 1/3$, the ground state is not a Laughlin state. However their study was not able to conclude any thermodynamic information about the state, and they speculated that  it may be superfluid. Our cluster mean-field study supports this claim by providing insight into the compressibility and the superfluid fraction. %and \textit{decreases} monotonically with increasing density. 
We now explore the structure of the metastable states seen in Fig.~\ref{eosplot}(a) (dashed). These  \textit{incompressible} phases (competing with the superfluid) are distinct from usual Mott insulators, as they occur at fractional densities.  They correspond to
filling factors $\nu = 1/2$ and $\nu = 1$, and are therefore suggestive of FQH physics. As we describe below, however, our solutions correspond commensurate density waves, rather than correlated liquids.  The energy difference between the metastable incompressible solution and the ground state at $n = 1/6$ and $1/3$ is very small, on the order $\Delta E \sim 0.1t$ within our calculations.

%Although it does not correspond to the ground state for $\phi=1/3$, it illustrates that there are allowed higher energy \textit{incompressible} phases (competing with the superfluid) at filling factors $\nu = 1/2$ and $\nu = 1$, which is suggestive of FQH physics. Moreover, these uncondensed phases are also distinct from usual Mott insulators, as they occur at fractional densities.  %Furthermore, these correspond to zero temperature, \textit{non-condensed} phases of bosons, that are not Mott insulators. 

\begin{figure}
\begin{picture}(100, 110)
\centering
\put(58, -10){\includegraphics[scale=0.275]{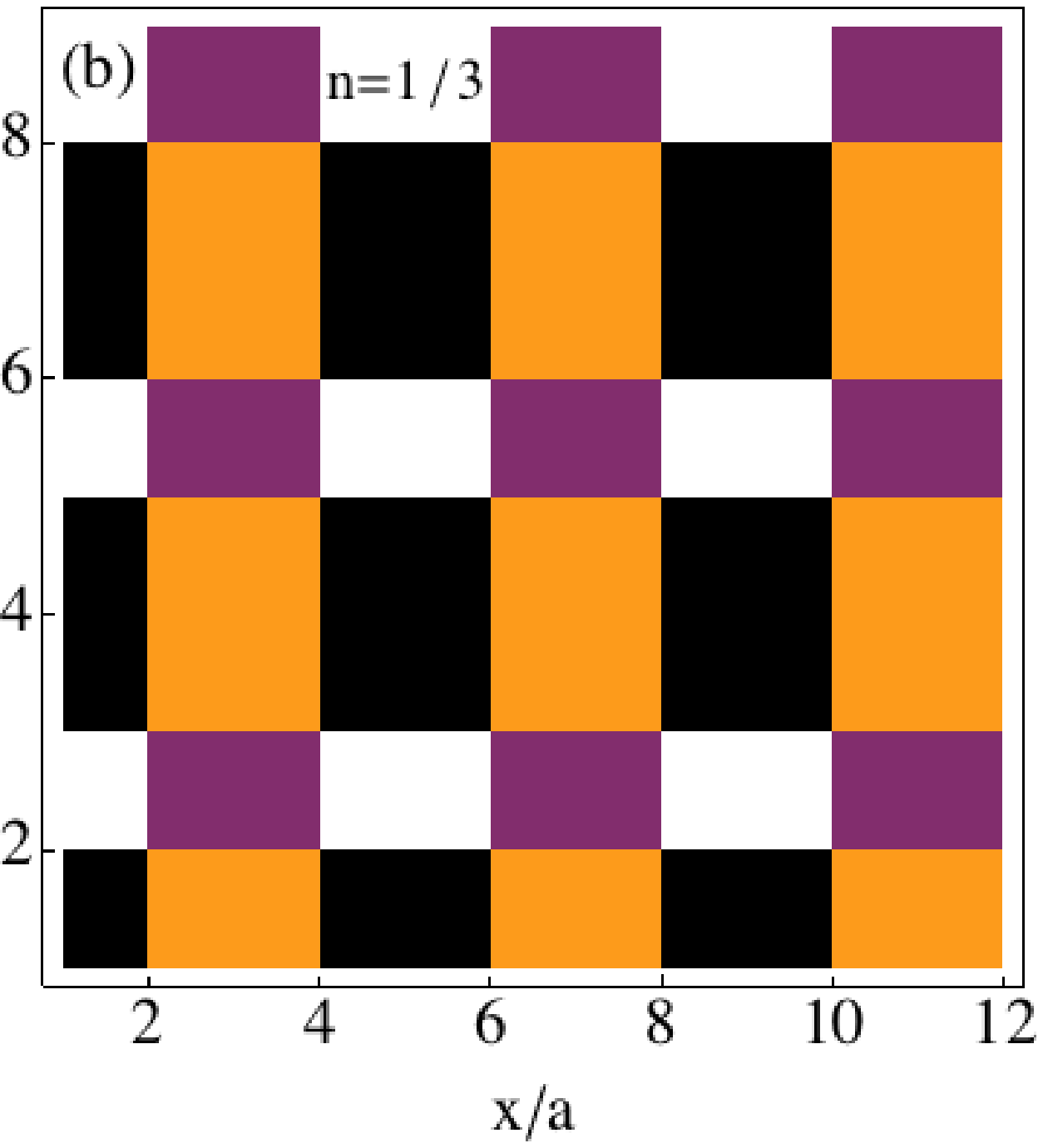}}
\put(160, 17){\includegraphics[scale=0.3]{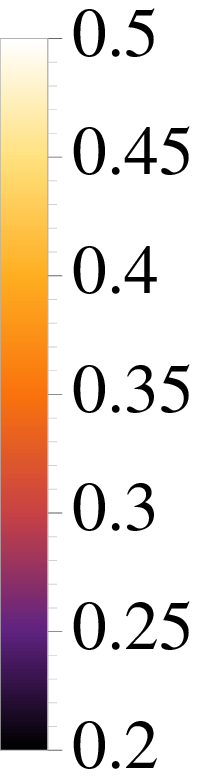}}
\put(-75, -10){\includegraphics[scale=0.3]{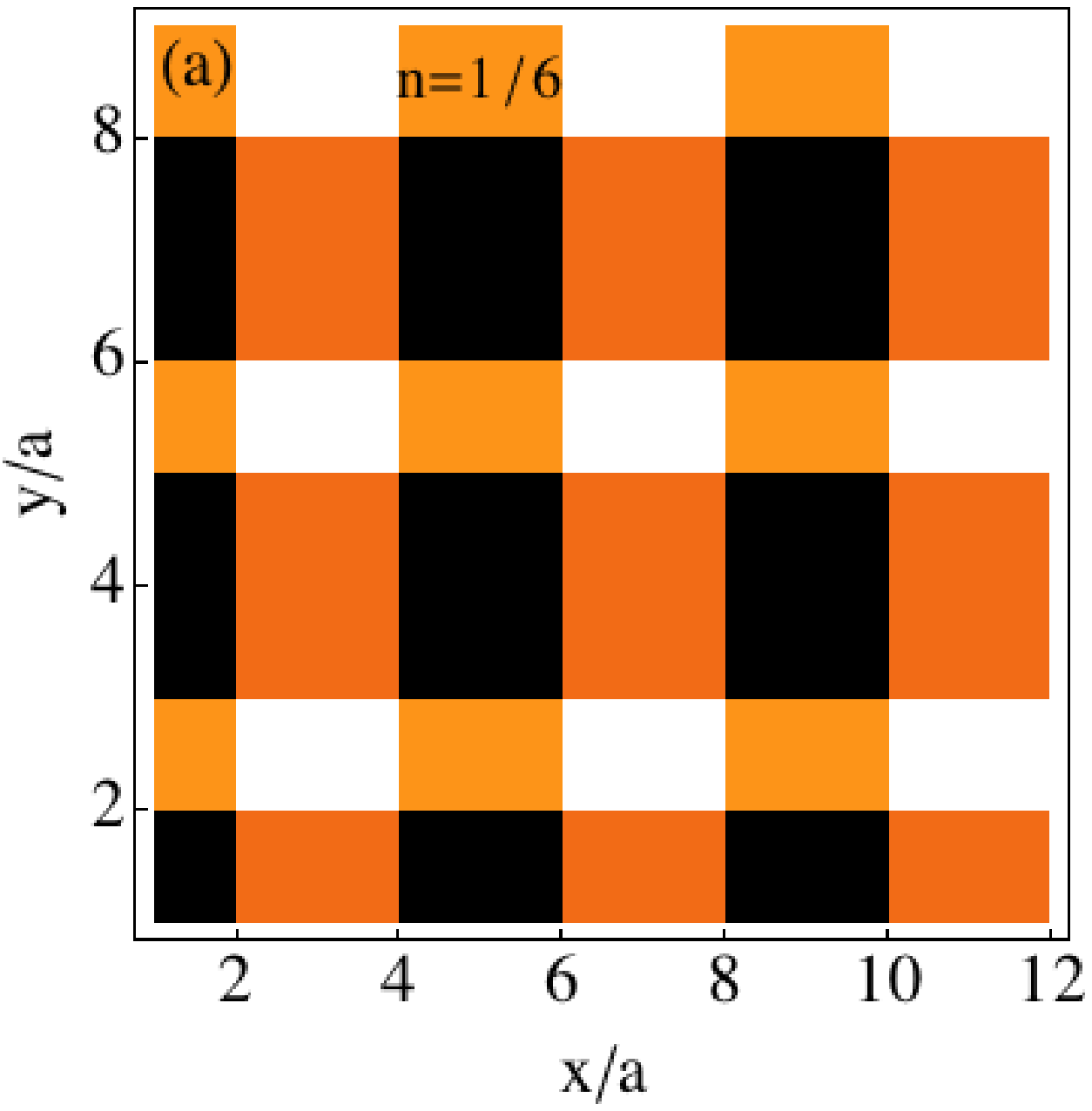}}
\put(35, 17){\includegraphics[scale=0.3]{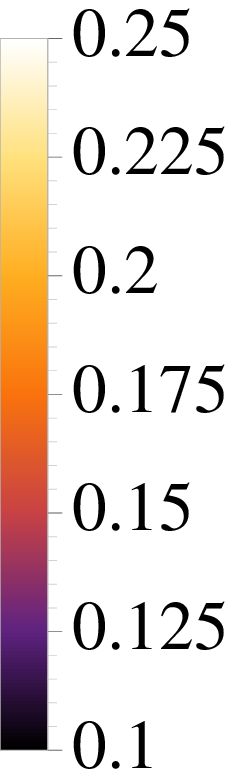}}
%\put(70, 55){\includegraphics[angle=270, origin=c,scale=0.39]{condoverdens.eps}}
\end{picture}
\caption{\label{cdws} (Color Online) Metastable checkerboard insulators at (a) $n = 1/6$ ($\nu = 1/2$) and (b) $n = 1/3$ ($\nu = 1$). These correspond to uncondensed phases which have higher energy than the corresponding superfluid ground state. The energy difference between the ground and metastable states is of order $0.1 t$.}
\end{figure}

To understand the physics of the incompressible states, we study a larger $12 \times 9$ system obtained by coupling $4 \times 3$ clusters using mean-fields. Using periodic boundary conditions, in Fig.~\ref{cdws}, we plot the density profile for the metastable states at $\tilde \mu = -1.7, -1$, corresponding to $\nu = 1/2$ and $1$ respectively.  In both cases, a strong checkerboard density wave order is apparent. Such commensurate density waves are natural in this system, and compete with superfluids and correlated liquids. We speculate that there are likely other metastable states corresponding to FQH liquids.  Longer range interactions or longer range hoppings can stabilize these FQH states \cite{Kapit10}, though our ansatz is not suited for finding them.

%In both cases, we find checkerboard density wave order. The appearance of an inhomogeneous density profile rules out the possibility of an incompressible liquid, like the $\nu= 1/2$ Laughlin state within our theory.  We cannot, however, rule out that this may be an artifact of our mean-field approximation. To see why, note that  as the boundary conditions are implemented via the mean-fields, these nearly zero mean-field states effectively correspond to exact solutions of the cluster with open boundary conditions. For the small clusters we consider, this can produce sizable finite size effects. Nonetheless, our finding that the ground state at $\phi = 1/3$ is \textit{not} an incompressible liquid is consistent with ED studies, which find very little overlap between the exact ground state and trial FQH wave-functions \cite{Sorensen05, Moller09}. At $\phi = 1/3$, the superfluid state has a uniform density. 

%Nonetheless, density-wave order is typically associated with long range physics, and have been predicted in rotating dipolar bosons \cite{Cooper05}, and studies of hardcore bosons and fermions in flat band models with long range parameters \cite{Sheng11, Kai15, Sheng15}. It is therefore remarkable that density wave order occurs in our model, with purely on-site interactions and a large magnetic field. At $\phi = 1/3$, the superfluid ground state we obtain is homogeneous. The conclusion of a superfluid ground state at $\phi = 1/3$ is consistent with ED studies, 
Our observation of checkerboard-like incompressible states highlights an important distinction between the continuum and lattice quantum Hall problem. In free space, translational symmetry breaking phases such as Wigner crystals, stripe, and bubble phases can compete with FQH liquids \cite{Fogler96, Fogler96-2, Anderson79, Fukuyama79, Lee83, Cooper05}. However these ``crystals" are distinguished from Laughlin states as they are almost always \textit{compressible}, and have gapless Goldstone modes, in contrast to the gapped FQH liquids.  In the lattice problem however, \textit{gapped} incompressible density-wave phases which break discrete lattice symmetries, such as the checkerboard phase, are possible, and may indeed be favored over more correlated incompressible liquids. 

\begin{figure}[h]
\begin{picture}(100, 190)
\centering
\put(-30, 110){\includegraphics[scale=0.45]{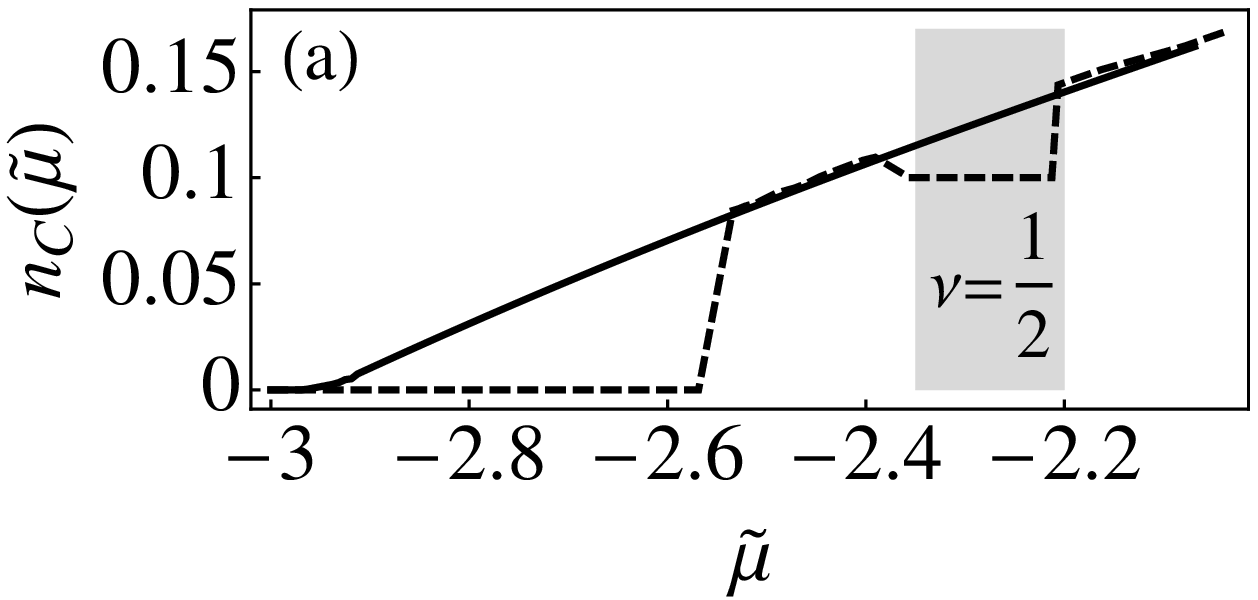}}
\put(-80, -10){\includegraphics[scale=0.32]{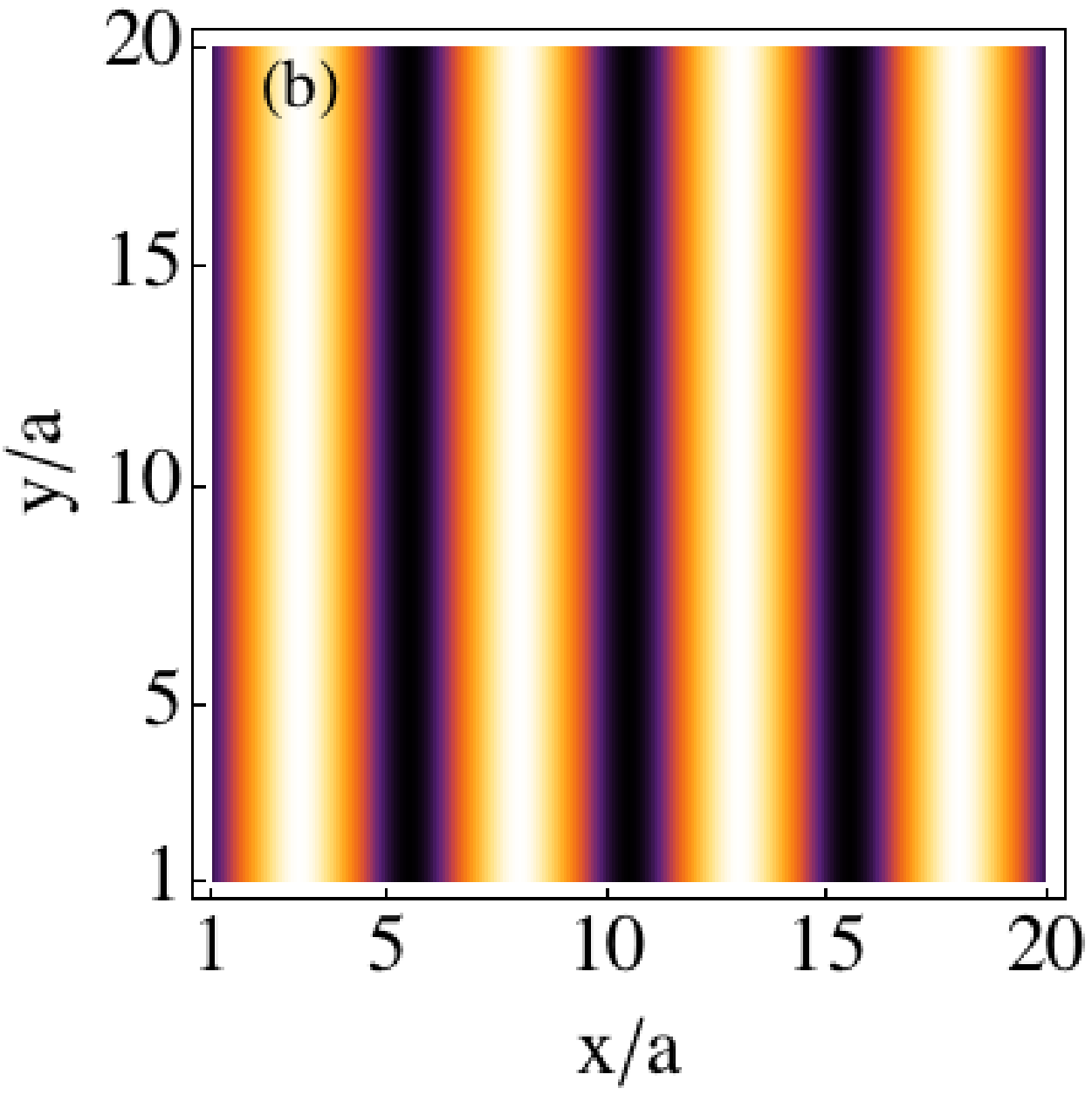}}
\put(41, -8){\includegraphics[scale=0.28]{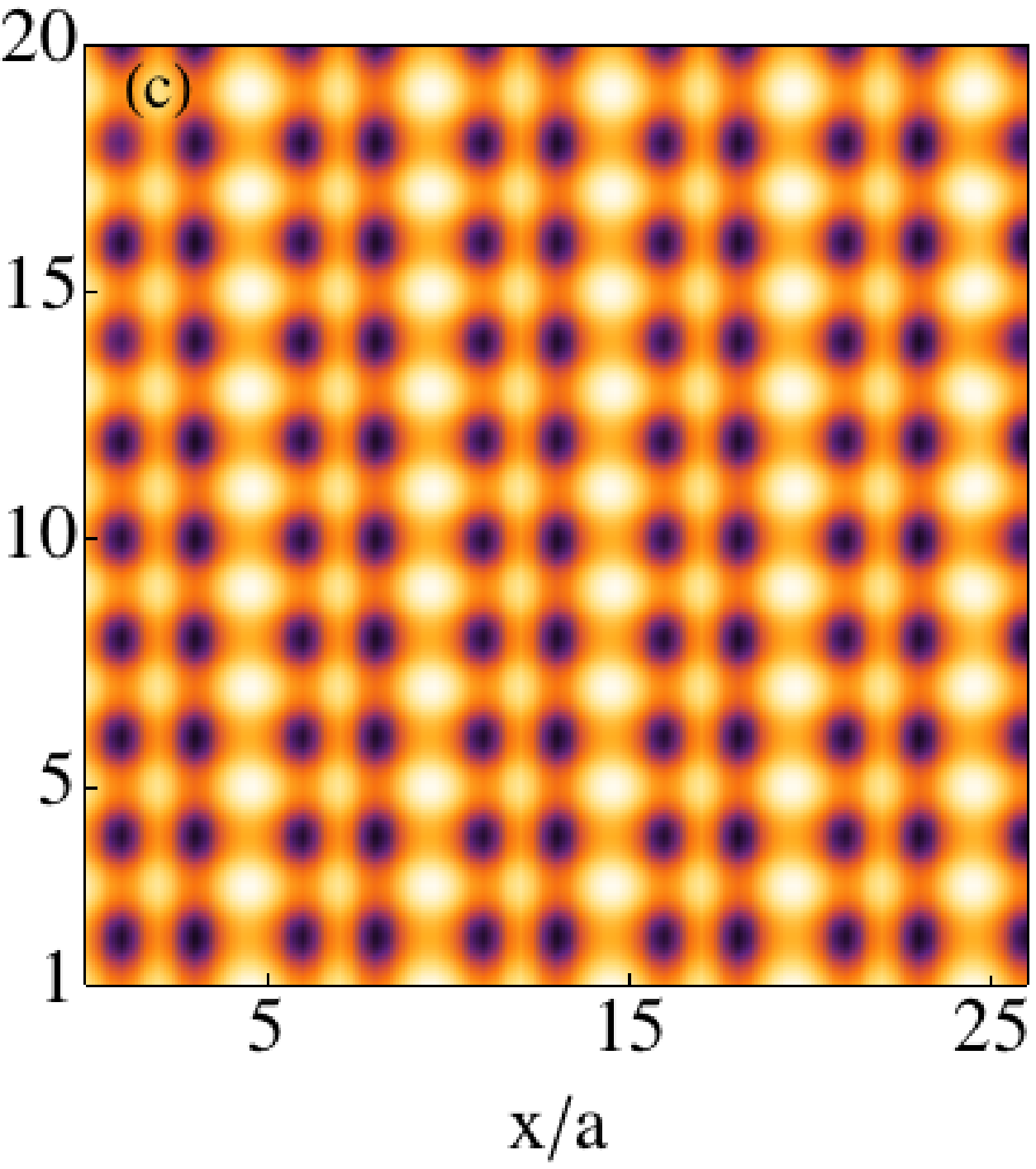}}
\put(150, 15){\includegraphics[scale=0.33]{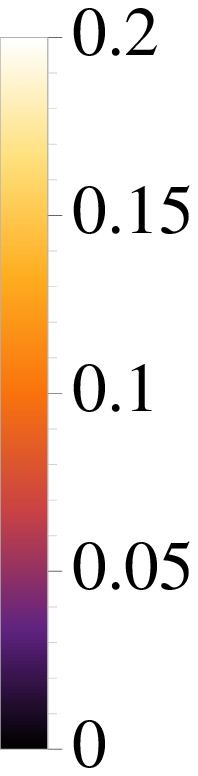}}
%\put(70, 55){\includegraphics[angle=270, origin=c,scale=0.39]{condoverdens.eps}}

\end{picture}
\caption{\label{supersolid} (a) (Color Online) Equation of state for $\phi = 1/5$ showing ground state (dashed) and metastable (solid). In this case, the incompressible plateau (dashed curve) at $\nu = 1/2$ or $n= 0.1$ corresponds to the ground state, while the superfluid has higher energy. (b) Density profile in the incompressible ground state at $n= 0.1$ showing stripe order. (c) The metastable superfluid phase at $n \approx 0.1$ also supports density wave order, and is analogous to a supersolid.}
%Density profile of the metastable compressible superfluid phase showing density wave order $\phi = 1/5$ and $n \approx 0.1$.}
\end{figure}

\subsection{Hardcore bosons at $\phi = 1/5$} 

Does the Harper-Hofstadter-Mott model support an uncondensed bosonic \textit{ground} state at fractional filling? Motivated by this question, we repeat our calculations for a smaller flux $\phi = 1/5$, which is closer to the continuum limit.

\begin{figure*}
\begin{picture}(100, 150)
\centering
\put(-210, 0){\includegraphics[scale=0.54]{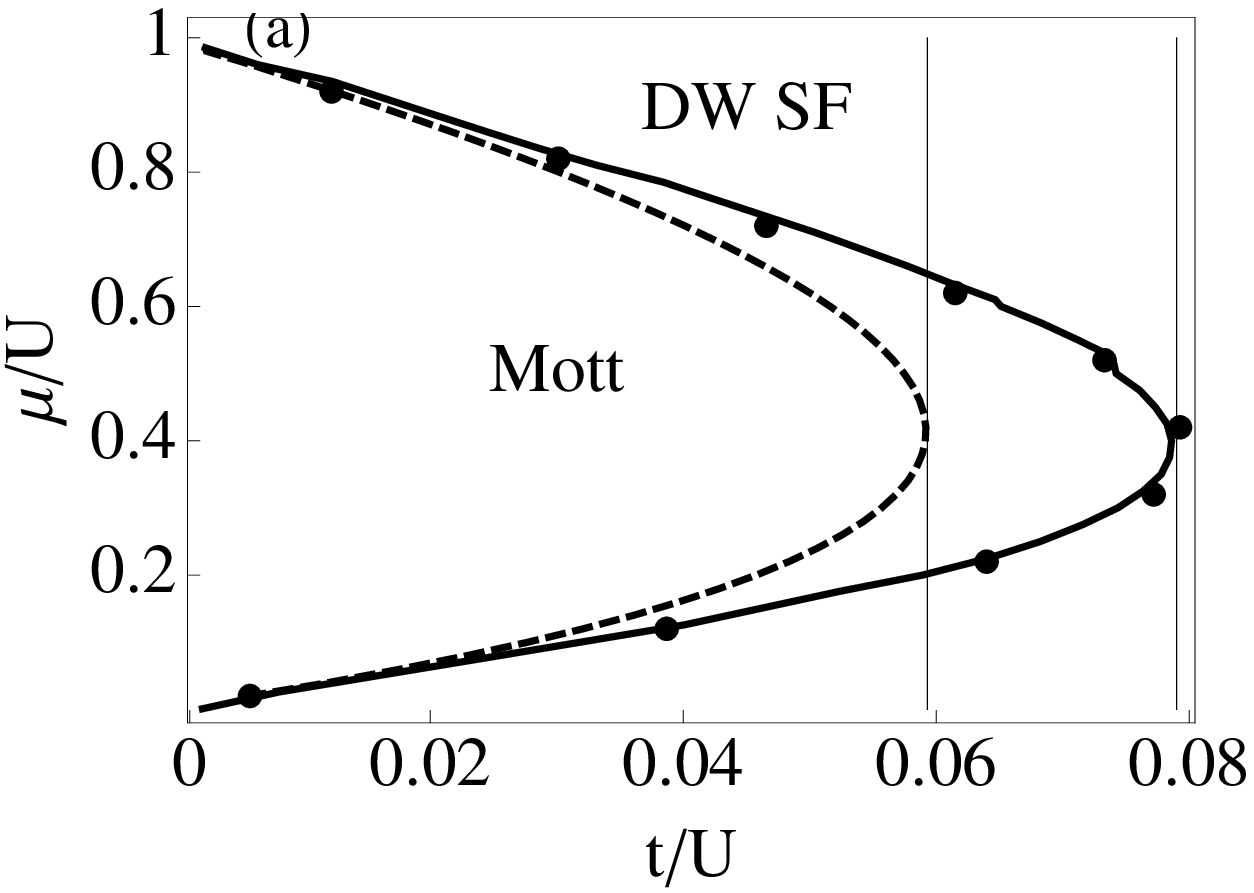}}
\put(-20, 0){\includegraphics[scale=0.45]{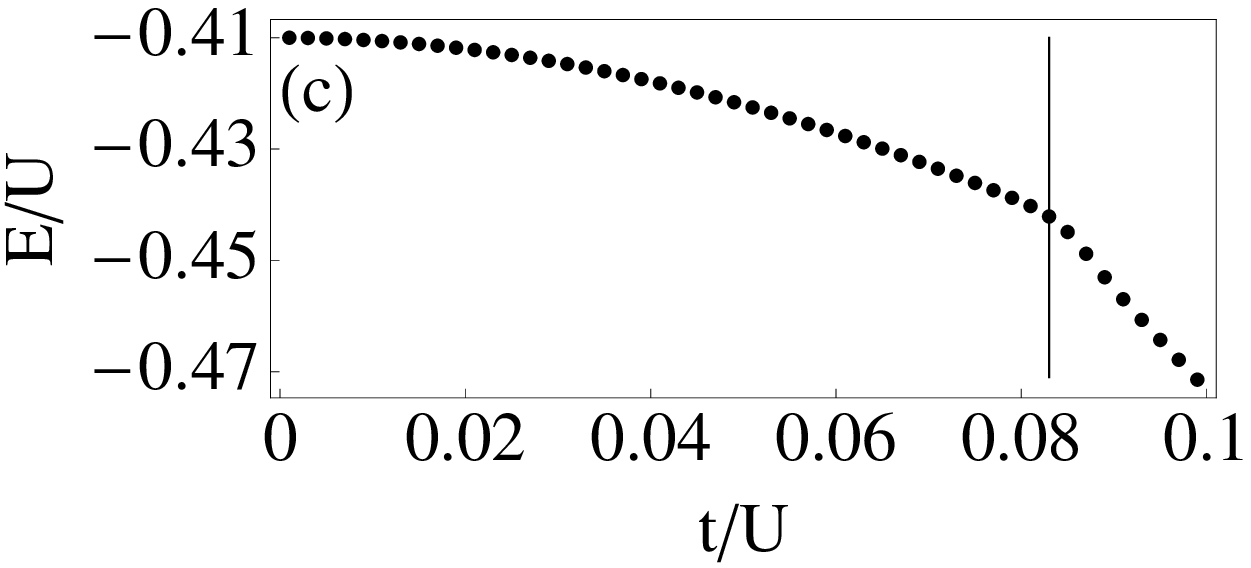}}
\put(-8, 90){\includegraphics[scale=0.405]{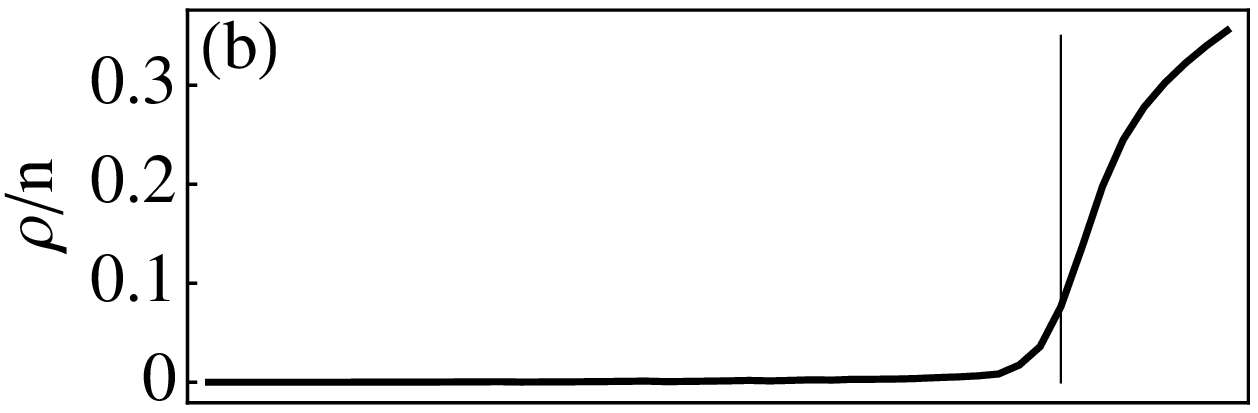}}
\put(145, -5){\includegraphics[scale=0.37]{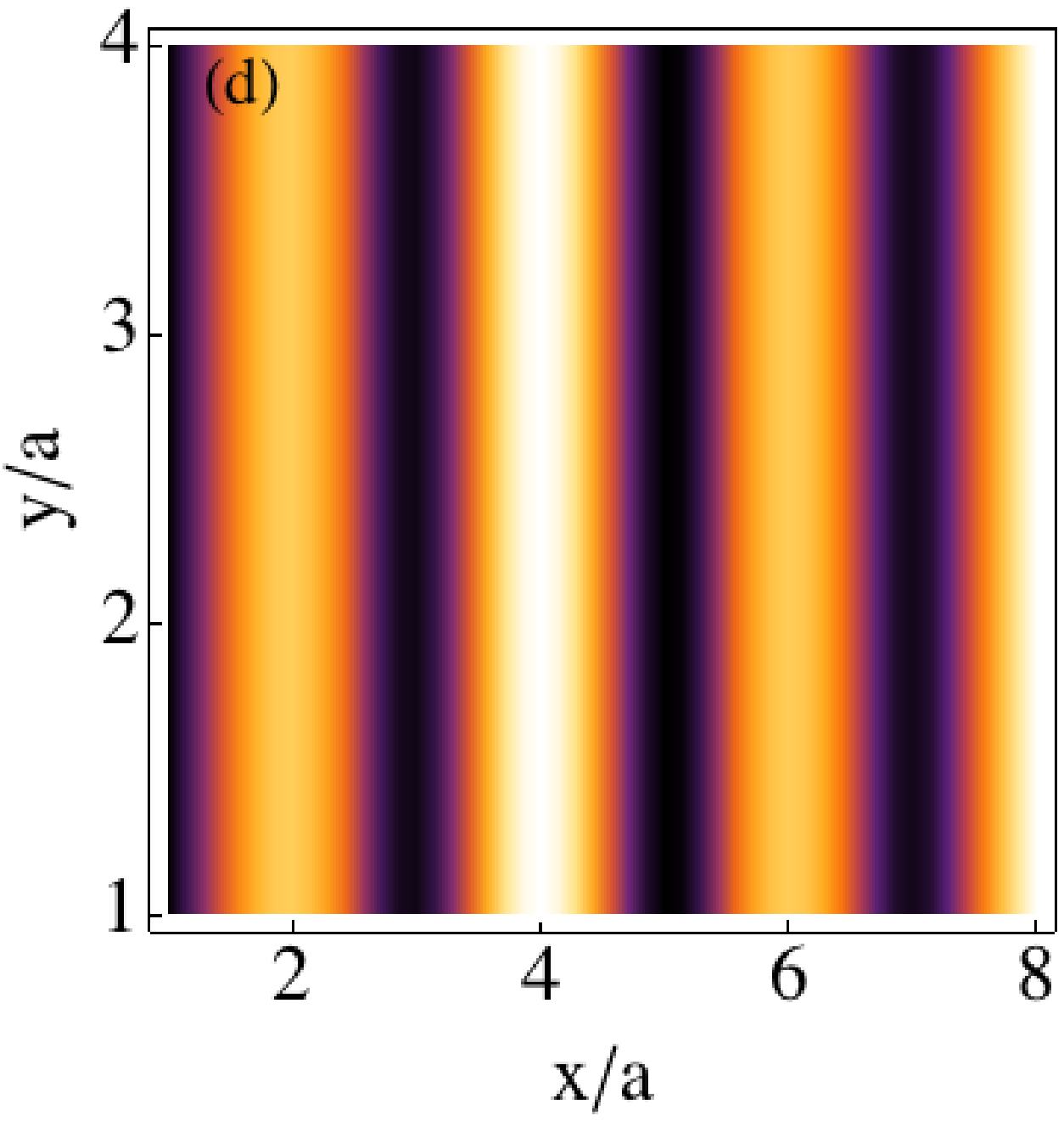}}
\put(285, 30){\includegraphics[scale=0.37]{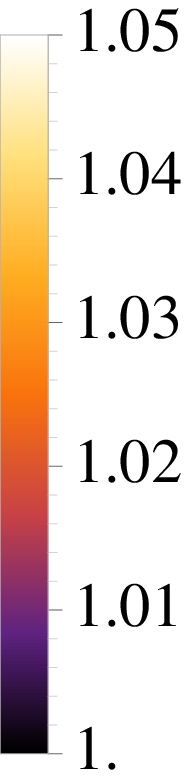}}

%\put(-68, -10){\includegraphics[scale=0.573]{condvsmu.eps}}
%\put(70, 55){\includegraphics[angle=270, origin=c,scale=0.39]{condoverdens.eps}}

\end{picture}
\caption{\label{pdfig} (Color Online) (a) Phase diagram of Harper-Hofstadter model at $\phi = 1/2$ using single site (dashed), $3\times 2$ (solid) and $4\times2$ (dots) clusters. Phase boundary separates $n=1$ Mott insulator and density-wave (DW) ordered superfluid (SF). Vertical line at $t/U \sim 0.06$ and $0.08$ mark the critical point $1\times1$ and $3\times2$ clusters respectively. The results of the $4\times2$ cluster are almost identical to the $3\times2$ results, except at the critical point. (b) Superfluid order parameter at $\tilde\mu = 0.4$ for a $4\times2$ cluster. (c) Mean-field energy per site for the same parameters as in (b). At small $t$, the energy scales quadratically as $t^{2}/U$, as predicted by Freericks and Monien in the absence of a magnetic field \cite{Freericks96}. The discontinuity in the slope of the energy versus $t$ marks the quantum phase transition to the superfluid. (d) Real space structure in the superfluid showing stripes on the order of few percent of the density. Much larger oscillations are found for the condensate order parameter (not shown).}
\end{figure*}

In Fig.~\ref{supersolid}(a), we plot the equation of state for a single $5 \times 2$ cluster at $\phi = 1/5$. Once again, all initial solutions converge to two distinct solutions: the dashed curve has \textit{lower} energy, indicative of an incompressible ground state of uncondensed bosons at $\nu = 1/2$. The solid curve is metastable, and is a superfluid with a somewhat larger superfluid fraction, which again decreases monotonically with density. At $n = 0.1$,  $\rho_{{\cal{C}}}/n_{{\cal{C}}} \approx 80\%$. Interestingly, as we show below, this weakly correlated superfluid has non-trivial real space structure.  

As previously described, we study larger systems by coupling clusters together with mean-fields. We plot the density of the resulting stable and metastable solutions at $n = 0.1$ ($\nu = 1/2$) in panels (b) and (c) respectively. As in the $\phi = 1/3$ case, the incompressible ground state shows density wave order, in particular, unidirectional stripe order. Here the geometry of the density waves is set by the cluster shape, and by changing the cluster, we can find different structures. For example, switching to a $2\times 5$ cluster switches the direction of the stripes. 

As before, we cannot reliably compare the energy of these density wave states and FQH liquids. Regardless, it is clear that for these parameters, the superfluid is not the ground state. This again is consistent with ED studies which find large overlap with the Laughlin wave function  at $\nu = 1/2$ \cite{Sorensen05} at these magnetic fields. The anisotropy of our clusters, and the corresponding appearance of stripe ordered solids may be related to the fact that in the thin-torus or Tao-Thouless limit \cite{Tao83}, Laughlin states are adiabatically connected to gapped density-wave solids \cite{Bergholtz05, Anderson83}, which might be what we are finding here.

We find that at $\phi = 1/5$, the superfluid phase also has density wave order \cite{Powell10}. This phase can be considered as the lattice analog of the supersolid phase in solid Helium \cite{Chan04}, or can be thought of as a high density vortex lattice. Superfluid models with density-wave order include rotating dipolar bosons \cite{Cooper05}, and studies of hardcore bosons and fermions in flat band models with long range parameters \cite{Sheng11, Kai15, Sheng15}. It is therefore remarkable that it occurs in our short-range model, with purely on-site interactions and nearest-neighbor hopping in an albeit large magnetic field \cite{Sheng11, Vasic15}.

\section{Physics near the vicinity of the Superfluid-Mott transition}

We now turn to the physics at higher densities, near unity occupation per site, specifically in the vicinity of the superfluid-Mott phase boundary for large, but finite $U$. At fixed flux $\phi$, we explore the parameter space spanned by $t/U$ and $\mu/U$. We focus on the $n= 1$ Mott lobe, and restrict our Hilbert space to $k =3$, or $0, 1, 2$ particles per site. Throughout, we use periodic boundary conditions on the entire $K \times L$ system. To facilitate comparison with ongoing experiments, we present results for $\phi = 1/2$ and $\phi = 1/4$, for clusters of size $3\times 2$ and $4\times 2$.

%explore a variety of cluster sizes and present results for $3\times 2$ and $4\times 2$ clusters at. %By exploring a wide range of chemical potentials \cite{natu-hof}, we can study several different filling fractions, unlike previous studies \cite{Onur09, Sorensen05, Hafezi07, Moller09, Moller15}, which tend to focus on a few particular choices of $\nu$, where FQH states can appear. 
\subsection{Superfluid-Mott phase boundary at $\phi = 1/2$}

We begin by discussing the $\phi = 1/2$ case, realized in the recent experiments of Miyake \textit{et al.} \cite{Miyake13} and Aidelsburger \textit{et al.} \cite{Aidelsburger13}. In Fig.~\ref{pdfig}(a) we present phase boundaries obtained for $1\times1$ (single-site Gutzwiller), $3\times2$ and $4\times 2$ clusters. For the $1\times1$ case, we reproduce the results of Umucalilar and Oktel Ref.~\cite{Onur07} at $\phi = 1/2$. We go beyond those results by exactly diagonalizing small clusters, finding that the phase boundaries are pushed outwards to larger tunneling. Panel (b) shows the mean value of the superfluid order parameter $\rho$ computed at $\mu/U = 0.4$ for a $4\times 2$ cluster as a function of $t/U$, which clearly reveals the second order nature of the phase transition. The phase boundaries are well converged already for modest cluster sizes $3\times2$; the $4\times2$ cluster boundaries are nearly identical to the $3\times2$ results, except very close to the tip of the Mott lobe. 

The larger Mott lobes in the cluster case are not surprising \cite{Luhmann13}: the single-site Gutzwiller method is known to overestimate the superfluid regions, as it only takes the hopping into account via mean-fields. Within the  Mott regions, the hopping is therefore identically zero, and the physics everywhere is identical to the ``atomic" ($t=0$) limit. In reality, particle or hole excitations created from the Mott insulating background can propagate and annihilate, which lowers the ground state energy of the Mott insulating region, enhancing its region of stability. In a square lattice, these corrections are found to be of order $t^{2}/U$ \cite{Freericks96}, and are captured by our numerics (Fig.~\ref{pdfig}(b)). The second-order superfluid to Mott transition is indicated by a kink in the energy versus  $t/U$, or a discontinuity in the slope, shown in Fig.~\ref{pdfig}(b). We use this to determine the phase boundaries presented in Fig.~\ref{pdfig}(a). To compare energies of competing phases, however, we must be careful in including the constant terms produced by the mean-fields of the form $t e^{i{\cal{A}}_{jk}}\langle a_{j} \rangle\langle a_{k} \rangle$. These terms do not affect the energies of the non-condensed solutions, but typically raise the energy of the superfluid. %As we show in the next section, for complex hopping, incompressible phases which have \textit{lower} energy than the superfluid appear \textit{outside} the Mott lobes. 

By coupling together clusters to form a larger system, we investigate the spatial structure of the superfluid and Mott lobes. As panel (d) shows, the superfluid state has density-wave order, analogous to a lattice supersolid. Near the Mott transition, the contrast in the density oscillations is rather small $\sim 5\%$. This is because $U \gg t$, and on-site interactions penalize density-wave order. We expect the contrast to grow as one approaches the weakly interacting limit \cite{Powell10}, where the fluid is more compressible. Even in this strongly interacting limit, the condensate fraction shows large oscillations of $\sim 20\%$. We caution however that although we believe the prediction of the density wave order to be qualitatively robust, and consistent with previous work by Powell \textit{et al.} \cite{Powell10}, the exact geometry of the oscillations depends on our cluster shape and the initial conditions. It is also worth noting that the supersolid phase we find breaks a discrete translational symmetry, and is distinct from the supersolid phase of solid He-$4$ \cite{Chan04}. 

The density wave order persists even in the Mott insulating state, however, the contrast in the oscillations is extremely small, as the Mott gap exponentially suppresses any density fluctuations. Within our mean-field theory, the superfluid-Mott transition is therefore associated with the restoration of a single, global $U(1)$ symmetry, and hence is second order. Full lattice translational symmetry is likely restored at a second critical point inside the Mott phase, but we cannot identify this phase transition, as there are no clear signatures in the energetics. A second possibility, is a direct \textit{first order} transition from a density-wave supersolid to a homogeneous Mott insulator \cite{Kuklov04}, but this is not supported by our numerics. 

\subsection{Competing Order at $\phi = 1/4$}

We now focus on the physics away from unity filling at $\phi = 1/4$, where previous studies have predicted the appearance of fractional quantum Hall states of excess particles skating on top of the Mott background \cite{Hafezi07, Onur07, Onur09}. An incompressible quantum Hall state would be characterized by a vanishing condensate fraction and a homogeneous density profile. We look for non-condensed solutions by using small trial values for the mean-fields, and iterate them to convergence.

\begin{figure}
\begin{picture}(100, 130)
\centering
\put(-60, -10){\includegraphics[scale=0.554]{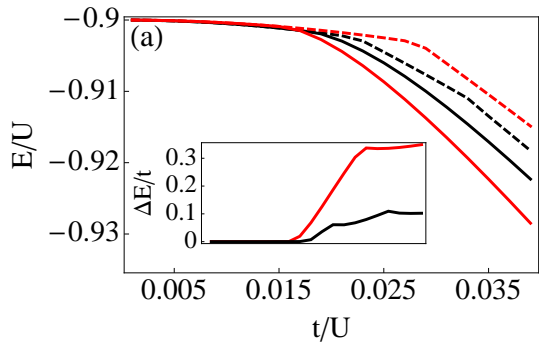}}
\end{picture}
\caption{\label{enfig} (Color Online) Mean-field energy for two competing mean-field solutions at $\phi = 1/4$ and $\tilde\mu = 0.9$ for $4\times2$ (Black) and $2\times2$ (red) clusters. Dashed lines are non-condensed, and correspond to states with integer total particle number, and may correspond to correlated states on top of the integer filling Mott insulator (see Fig.~\ref{densfig}). Solid lines indicate superfluid solutions which have lower energy. Inset shows energy difference between solid and dashed curves for $2\times 2$ (red) and $4\times2$(black) clusters for the same range of chemical potentials. Energy difference between superfluid and non-condensed states reduces rapidly with increasing cluster size.}
\end{figure}

\begin{figure}[h]
\begin{picture}(100, 150)
\centering
\put(-73, 80){\includegraphics[scale=0.557]{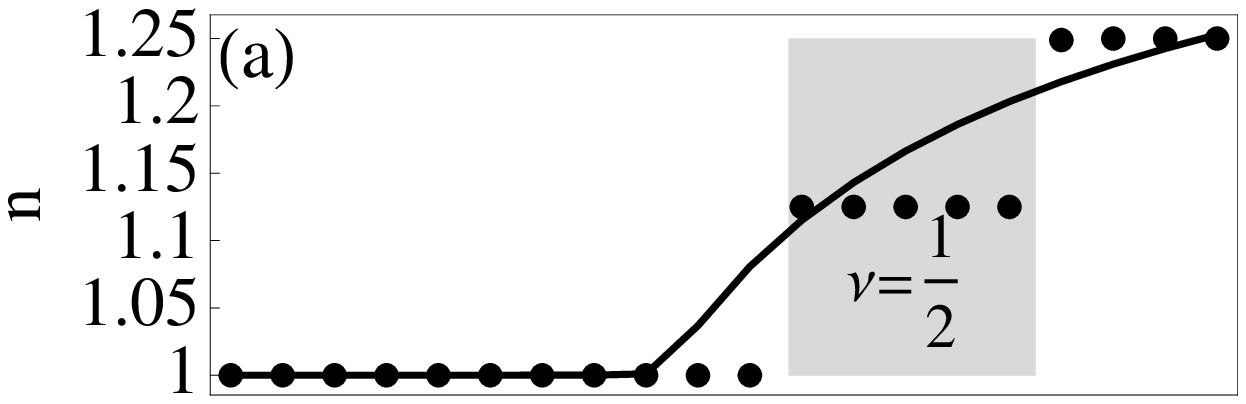}}
\put(-67, -10){\includegraphics[scale=0.545]{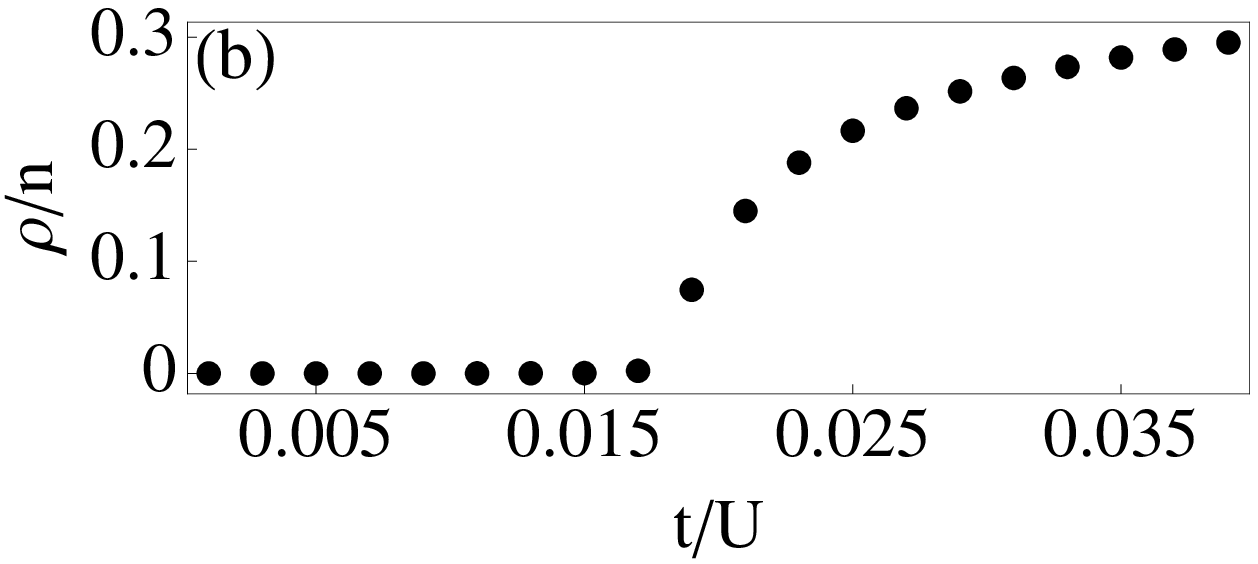}}
\end{picture}

\caption{\label{densfig} (Top) Density as a function of $\tilde t$ for a $4\times2$ cluster at $\tilde\mu= 0.9$. Solid curve shows the superfluid solution and dashed curve corresponds to non-condensed solutions with integer total particle number, or density $n = 1, 1+\epsilon, 1+2\epsilon$, where $\epsilon = 1/8 = 0.125$. The superfluid always has lower energy than the non-condensed solutions but the energy difference decreases rapidly with cluster size. (Bottom) Condensate fraction $\rho/n$ as a function of $t/U$ for the superfluid solution.}
\end{figure}

In Fig.~\ref{enfig}, we compare the mean-field energies for two solutions of the mean-field equations corresponding to zero and non-zero condensate order parameter at a fixed chemical potential $\tilde\mu = 0.9$. The black curves are drawn for $4\times2$ clusters whereas the red curves for $2\times2$ clusters. A similar picture holds for other values of $\tilde\mu$ sufficiently close to $1$.

In addition to the unity filling Mott insulator, where both solid and dashed curves (of a given color) yield identical results, our theory finds competing phases away from commensurate filling. In Fig.~\ref{densfig}(a), we plot the density for the same parameters as in Fig.~\ref{enfig} for the $4\times2$ cluster. The solid curve reveals a second-order phase transition to a superfluid phase, characterized by a non-zero superfluid order parameter. The dots show transitions between various non-condensed states $n = 1.125$ and $n = 1.25$, corresponding to $\nu = 1/2, 1$ respectively (see Fig.~\ref{densfig}). The $\nu= 1/2$ state at $\phi = 1/4$ is absent in $1\times1$ and  $2\times2$ clusters (not shown), and only appears for sufficiently large cluster sizes. As the boundary conditions are implemented via the mean-fields, these zero mean-field states correspond to exact solutions of the cluster with open boundary conditions. The Hamiltonian then commutes with total particle number, therefore they have an integer number of particles. The state at $n= 1.125 (1.25)$ corresponds to $1 (2)$ excess particles on top of the Mott background, as predicted by Refs.~\cite{Hafezi07, Onur07, Onur09}. Importantly the width of the non-condensed state at $\nu = 1/2$ is consistent with variational calculations of Umucalilar and Mueller \cite{Onur07}, who argue that this state is a $\nu  = 1/2$ Laughlin state. %Within our theory, these state is not a Laughlin state, although this may be due to finite size effects from the small cluster sizes we consider. 

Our cluster calculations however provides much better variational energies for the superfluid phase, as compared to single site mean-field theory \cite{Onur07, Onur09}. Comparing the energies of the two mean-field solutions in Fig.~\ref{enfig}, we find that the superfluid (solid red and black curves) \textit{always} has lower energy than the non-condensed states (dashed red and black curves). Interestingly however, the energy difference between the superfluid and the non-condensed states at fixed tunneling reduces rapidly with increasing cluster size. Indeed for $4\times2$ clusters, the largest energy difference (inset) is of order $0.1t$. For typical experimental parameters, this energy scale corresponds to a temperature on the order of $100$pK, which may be beyond the current experimental capability. Undoubtedly, the exact ground state  will be sensitive to experimental details such as nearest neighbor interactions, and the exact band dispersion. The issue of the actual ground state in the large system limit therefore remains an open question in this context except that our work establishes the presence of competing compressible and incompressible quantum phases close by in energies.

\section{Experimental Signatures}

Our work has important consequences for the experiments at MIT and Munich \cite{Miyake13, Aidelsburger13}, as the equation of state can be directly measured in trapped ultra-cold gases \cite{Nascimbene10}. In a harmonic trap, the compressibility can be measured by studying the equation of state, the \textit{in situ} density versus the chemical potential. The in-trap density can be measured either using high resolution imaging \cite{Gemelke09} or phase-contrast imaging \cite{Ryoung12}. 

The phase coherence in the strongly interacting superfluid and lattice supersolid phases can be readily probed using time-of-flight. For $\phi = 1/2$, the superfluid order parameter oscillates, leading to a larger unit cell, and additional peaks in the momentum distribution of the atoms. This structure is directly seen in time-of-flight images.

The incompressible phases are particularly striking in \textit{in situ} images, as they correspond to plateaus.  The density modulations in these phases may be observed using high resolution imaging \cite{bakr10, Sherson10}, light scattering, or Bragg spectroscopy \cite{Ketterle99}.

%A signature of the incompressible, crystalline phases is the vanishing condensate order parameter or a flat momentum distribution at fractional filling, which distinguishes them from usual Mott insulators. Single site resolved imaging techniques \cite{bakr10, Sherson10} or Bragg spectroscopy \cite{Ketterle99} can be used to detect density wave order in the crystalline and supersolid phases. 

In the absence of a magnetic field, the phase boundary between the superfluid and Mott insulator can be experimentally determined using a variety of local and global probes such as single site imaging \cite{Sherson10, bakr10}, measurements of thermodynamic variables such as the compressibility \cite{Gemelke09},  time-of-flight \cite{Greiner02, Garcia10} and band mapping \cite{natu12}. All of these techniques can be extended to the case of non-zero magnetic field. The determination of the phase boundary and observing its deviation from the mean-field results is one of the first steps to exploring correlation effects in the vicinity of Mott lobes in the Harper-Hofstadter-Mott model.  %For $\phi = 1/2$, The density oscillations, although small, may appear in high resolution imaging or Bragg spectroscopy \cite{Ketterle99}. %Compressibility ($\kappa^{-1} = \partial n/\partial\mu$) is the most direct probe of a non-condensed state away from unity filling. 

Temperature will be the key challenge in observing incompressible phases away from integer filling since the energetics here distinguishing different competing phases are very small in general.. As our study shows, the energy difference between competing phases away from commensurate filling can be extremely small, on the order of $0.1t$. This is beyond the current reach of most experiments, especially as the Raman lasers lead to significant heating from spontaneous emission. One approach to mitigate heating is to use the spin degree of freedom as a ``synthetic" dimension \cite{Stuhl15, Fallani15}; in this case, the strength of the Raman lasers only has to be on the order of the energy to flip a spin (which corresponds to a hopping process in the synthetic dimension), which is typically much smaller than the lattice recoil energy $E_{R}  = \hbar k^{2}_{R}/2m$. However as the density-density and spin-spin interactions become ``long ranged" in this synthetic dimension language, the underlying Hamiltonian and band structure is qualitatively different from the Harper-Hofstadter-Mott model we study. Whether bosonic  FQH phases can even occur in these long-ranged lattice systems is an open question. 

\section{Conclusions}

To conclude, we have presented an efficient numerical technique for exploring strongly correlated bosonic systems, such as lattice bosons in a strong magnetic field. We have therefore theoretically explored by a cluster-mean-field technique the competing quantum many-body ground states of the Harper-Hofstadter-Mott model. By exactly diagonalizing small clusters, our approach includes local quantum correlations typically absent in purely mean-field theories. Our approach captures more global features by coupling neighboring clusters using mean-fields (Although less exact than ED, our work allows for studying systems larger than what typical exact diagonalization or Quantum Monte Carlo can accomplish). This allows us to obtain improved variational energies for superfluid and Mott states, and explore other exotic states. 

We presented the equation of state of a strongly interacting, dilute Bose gas in a strong magnetic field as a function of chemical potential for small and large values of flux. For less than one particle per site, at filling fractions away from $\nu = 1/2$ and $1$, the ground state is always superfluid. The superfluid fraction
decreases monotonically with increasing density; at $\nu = 1/2$, the superfluid fraction is only $\sim 30\%$, indicative of a strongly correlated state, analogous to He-4. At $\nu = 1/2$ and $1$, incompressible density wave states appear, but these are metastable for large flux. These states are found to be density wave checkerboard solids. We do not find any evidence for a Laughlin ground state within our calculations. Our not finding a fractional quantum Hall type incompressible ground state in this context is consistent with the fact that even in (the continuum) two-dimensional electron systems, the fractional quantum Hall states have very fragile energetics and are typically found to be the ground state of the system only in the lowest orbital Landau level (and not for very small filling factors, typically above 1/7 filling) and essentially never in the third Landau level (with the second Landau level being marginal where the fractional quantum Hall states compete with density wave type states).  We cannot of course rule out the existence of possible fractional quantum Hall states in our lattice system in the thermodynamic limit (because of the very small energy differences we find differentiating the different possible many-body ground states), but it is clear that superfluid and incompressible density wave states are typically the generically preferred states in the lattice Harper-Hofstadter-Mott system.

We then extended our results to unity filling by going beyond the hard core limit. At $\phi =1/2$, we obtained the superfluid-Mott phase boundary in two dimensions. The phase transition is found to be continuous, and is characterized by a vanishing condensate order parameter. Both the superfluid and the Mott state have density-wave order, although the contrast of the density oscillations in the Mott state are extremely small. In the superfluid phase, density oscillations are accompanied by a spatially oscillating order parameter, which could be readily observed in time-of-flight or band-mapping measurements.  

At $\phi = 1/4$, we obtain correlated, non-condensed states away from commensurate filling, at densities of $n= 1.125$ and $1.25$, corresponding to $\nu = 1/2$ and $1$. These states are absent in the usual single site mean-field theory \cite{Onur07}, but appear for sufficiently large cluster sizes, such as the $4\times2$ cluster we consider. We find that although the ground state away from commensurate filling is a superfluid, the energy difference between the superfluid ground state and the non-condensed metastable states decreases with increasing cluster size. The exact nature of the ground state in the thermodynamic limit is uncertain.

Finally, we remark that the advantage of our cluster method over sophisticated techniques such as DMRG or ED is its simplicity. It is versatile, and efficient even in dimensions greater than two \cite{Luhmann13}, and can be readily generalized to include long range hopping, interactions, other lattice geometries. We therefore expect this method to complement very well ongoing ED studies of the integer and quantum Hall effects in topological flat band systems.

%can be extended to models with longer range hoppings and interactions with and without a magnetic field, or lattice geometries which host flat bands, to explore bosonic Chern insulators \cite{Moller15, Kapit10, Regnault11, Sheng11}. Finally, we remark that our numerical technique can complement ongoing studies of lattice bosons in low dimensional systems in large magnetic fields using DMRG and variational techniques \cite{Atala14, Petrescu13, Piraud15, Petrescu15, Moller09}. 

%The cluster technique complements other numerical techniques, and offers a novel way to explore competing order such as strongly correlated superfluids, supersolids, and insulators at fractional filling.

%Furthermore, it has the advantage that it 

\section{Acknowledgements} SN and SDS would like to thank the LPS-MPO-CMTC, JQI-NSF-PFC and ARO-Atomtronics-MURI for support. SN is grateful to the NSF through the PFC seed grant ``Emergent phenomena in interacting spin-orbit coupled gases" for support. EJM acknowledges support from NSF 1508300. SN would also like to thank Srinivas Raghu, Kai Sun and Kaden R. A. Hazzard for stimulating discussions, and Stanford and Cornell Universities for their hospitality during the completion of this work. 

\bibliography{hofstadterbib}

\end{document}